**International Journal of Artificial Intelligence In Medicine (IJAIMED)**
Volume 1, Issue 01, Jan-Dec 2023, pp. 1-8. Article ID: IJAIMED_01_01_001
Available online at https://iaeme.com/Home/issue/IJAIMED?Volume=1&Issue=1
**Journal ID:** 2227-1989, **DOI:** https://doi.org/10.17605/OSF.IO/35N7E
© IAEME Publication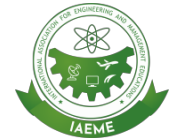

# SEGMENT ANYTHING MODEL (SAM) FOR BRAIN EXTRACTION IN fMRI STUDIES

**Dwith Chenna**

University of Maryland, College Park, MD, United States

**Suyash Bhogawar**

George Washington University, Washington DC, United States**ABSTRACT**

*Brain extraction and removal of skull artifacts from magnetic resonance images (MRI) is an important preprocessing step in neuroimaging analysis. There are many tools developed to handle human fMRI images, which could involve manual steps for verifying results from brain segmentation that makes it time consuming and inefficient. In this study, we will use the segment anything model (SAM), a freely available neural network released by Meta[4], which has shown promising results in many generic segmentation applications. We will analyze the efficiency of SAM for neuroimaging brain segmentation by removing skull artifacts. The results of the experiments showed promising results that explore using automated segmentation algorithms for neuroimaging without the need to train on custom medical imaging dataset.*

**Keywords:** Segment Anything Model (SAM), fMRI, Segmentation, Neuroimaging, Brain extraction.**Cite this Article:** Dwith Chenna and Suyash Bhogawar, Segment Anything Model (SAM) For Brain Extraction in fMRI Studies, International Journal of Artificial Intelligence in Medicine (IJAIMED), 1(1), 2023, pp. 1-8.
https://iaeme.com/Home/issue/IJAIMED?Volume=1&Issue=1

## 1. INTRODUCTION

Magnetic resonance imaging (MRI) has made available high resolution 3D imaging data of the human brain, which are instrumental in the field of neuroimaging. These MRIs can be used to diagnose and monitor several brain disorders like Alzheimer's disease, schizophrenia, epilepsy, multiple sclerosis and depression. A wide variety of such medical research into these disorders involves segmentation as one of the preliminary steps in medical image processing. Especially accurate segmentation of MRI images is essential for diagnosis, monitoring and treatment of a wide variety of such neurological disorders. Manual segmentation is considered the best method but is laborious, time consuming and doesn't scale for large neuroimaging studies. Many researchers have attempted to address this with automatic segmentation, to make it scalable for large datasets.

https://iaeme.com/Home/journal/IJAIMED 1 editor@iaeme.com

Many software tools for neuroimaging studies that automate the segmentation of brain region and structure like (1) FreeSurfer, (2) FMRIB Software Library (FSL) and (3) Statistical Parametric Mapping (SPM). While these tools are not just limited to segmentation, it is an important step in the preprocessing of medical imaging data. Several studies have been made to judge the relative comparison of these publicly available software tools and algorithms [1-2,18].

Machine learning approaches based on deep learning are also being widely used for robust segmentation of medical images [3]. These models were mainly limited by the dataset used for training and had limited ability to generalize for real use cases. MRI images from magnetic resonance (MR) scanners require multiple preprocessing including segmentation. These pre-processing steps are used in many of the widely used tools for neuroimaging.

In this paper we will analyze the segmentation algorithm for brain extraction, which is an essential preprocessing step for neuroimaging. We analyze the accuracy of the Segment Anything Model (SAM), which is the latest general segmentation network released by Meta[4], for brain extraction. The accuracy of the model could indicate efficiency of robust segmentation algorithms for brain extraction, without the restrictions on availability of diverse medical data.

## 2. BRAIN EXTRACTION

Brain MRI scan provides a 3D image of the brain scanned in x, y, z direction, usually with 1 to 2 mm thickness of slice in each direction. The quality of the scan depends on MR scanner, slice thickness, scan time, scanning sequence and protocol. There are multiple sources of noise including head movement . The brain structure is also popularly represented as 2D images in (x, y), (y, z) and (x, z) planes. The removal of non brain tissues such as skull, neck, muscles, bones, fat is referred to as brain extraction. Many of the popular tools used probabilistic methods [5], inflating sphere until brain boundary [6], atlas based label propagation [7] to name a few methods for brain extraction. We employ a segmentation algorithm, to classify 2D image pixels as brain or non-brain pixels. The brain region is extracted using a mask for extraction of the brain region. These 2D image presentations can be used by segmentation algorithms.

## 3. SEGMENTATION ALGORITHM

Many deep learning based CNN models have been developed for automatic segmentation of brain regions, especially the anatomical structures like hippocampus and cerebellum. These models might not generalize well for the variations in the complexities of medical imaging, especially with the limited data due to privacy reasons. Larger models with an increased number of layers have shown promising results in neuroimaging dataset [8]. These CNN architectures are made of multiple layers consisting of (i) input layers processing raw input data (ii) convolution layers with filters to map different feature maps (iii) activation functions to introduce non-linearity to the models, which helps with better learning/representational capabilties (iv) pooling layers to down sample the features and extract relevant information. U-Net is a popular architecture used for medical imaging, which uses cross layer connection for better segmentation results. In this section we review the architecture for the Segment Anything Model (SAM) recently released by Meta [4].

The three main components of SAM are i) image encoder ii) prompt encoder and iii) fast mask decoder.

*Image encoder*

The image encoder model is based on Vision Transformer (ViT) [9], which is designed for real-time performance on high resolution inputs. The pre-trained ViT model uses ViT-H/16 with 14x14 window attention and four equally spaced global attention blocks [10]. The output of the encoded models is downscaled input image, which is computed only once for the image and can be reused with different prompts. The architecture uses efficient computation by using 1x1 convolution to get 256 channels followed by 3x3 convolution followed by layer normalization. Input image resolution is 1024x1024 either through scaling or padding, which results in 64x64 embeddings.

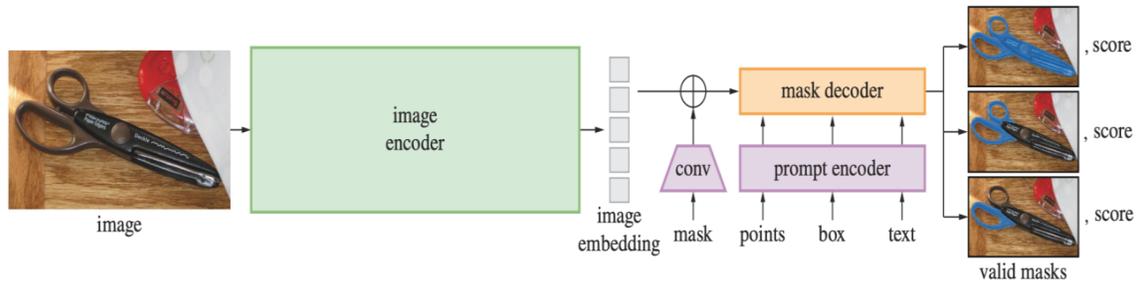

**Fig 1.** Segment Anything Model (SAM) architecture []

*Prompt encoder*

The prompts take as input points, boxes, text or dense masks, which are represented through embedding using different types of encoding. For points the positional encoding represents points location and learned embedding to indicate foreground or background. Box is represented through two positional encoding representing top-left corner and bottom right corner, with learned embedding. Text uses a text encode from CLIP [11], dense prompts are summed after generating embedding using convolutions. The prompts are encoded into 256 size vector embedding. Dense prompts have spatial correspondence with input image, the resolution of the masks is at 4x lower resolution, then downscale it additional 4x using stride 2 convolution with 4, 16 output channels and finally with 1x1 convolutions that maps to 256 vectors. These mask embedding are added elementwise, if there is no mask prompt then a learned no-mask embedding is used.

*Mask decoder*

Mask decoder is used to map the image embedding, prompt embedding and output token to output masks. The mask decoder architecture [12-13], uses a modified Transformer decoder block [14] with a dynamic mask prediction head. The decode uses bidirectional self-attenttention and cross-attention via prompt-to-image embedding and vice-versa. The mask decoder layer does the following steps: (1) self-attention on the tokens (2) cross-attention from tokens to the image embedding (3) point-wise MLP updates each token (4) cross-attention from the image embeddings to tokens. The updated tokens and image embeddings from the previous layers. Finally a dynamic linear classifier to compute mask probabilities for each location.

## 4. IMPLEMENTATION

In this section, we will discuss the implementation details of using SAM for fMRI brain extraction. The dataset used for this study is from OpenNuro [15], open source platform to share BIDS[17] compliant medical imaging data, OpenNeuro Dataset ds004590 [16].

## 4.1. Results

The algorithm is validated on sagittal, coronal and axial image slices generated from four different subjects in 4-5 different sessions. Fig 2 shows sample images for Sagittal, Coronal and Axial slices for the 3D MRI scan.

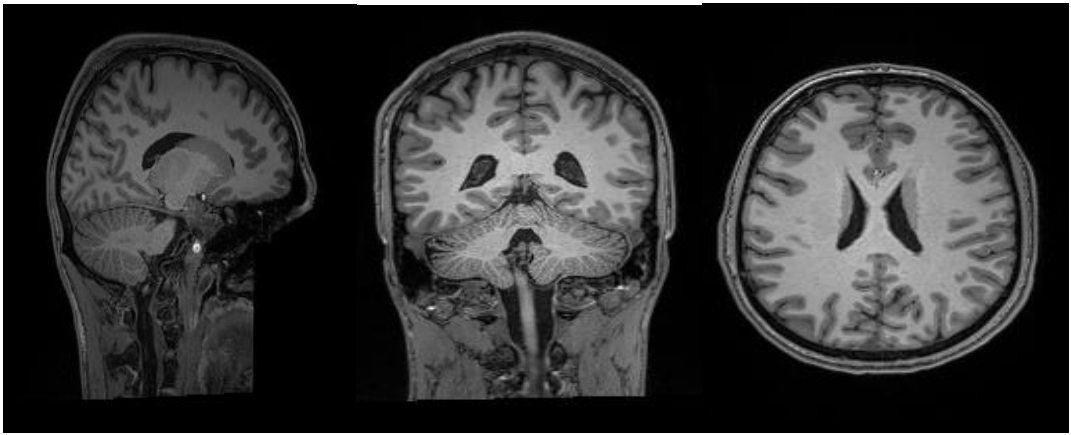

**(a)** **(b)** **(c)**

**Fig 2.** Sample image from dataset with (a) Sagittal, (b) Cornal and (c)Axial slices

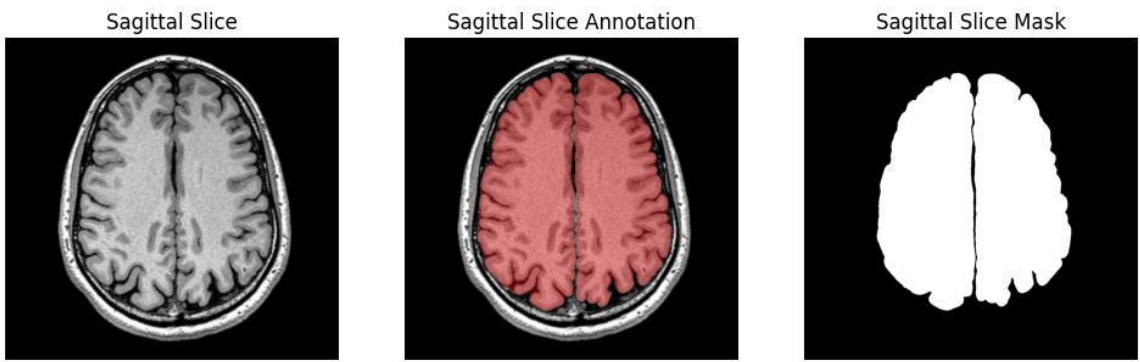

**Fig 3.** Results of brain extraction (a) MRI sagittal slice (b) brain segment annotation (c) brain segmentation mask generated using SAM

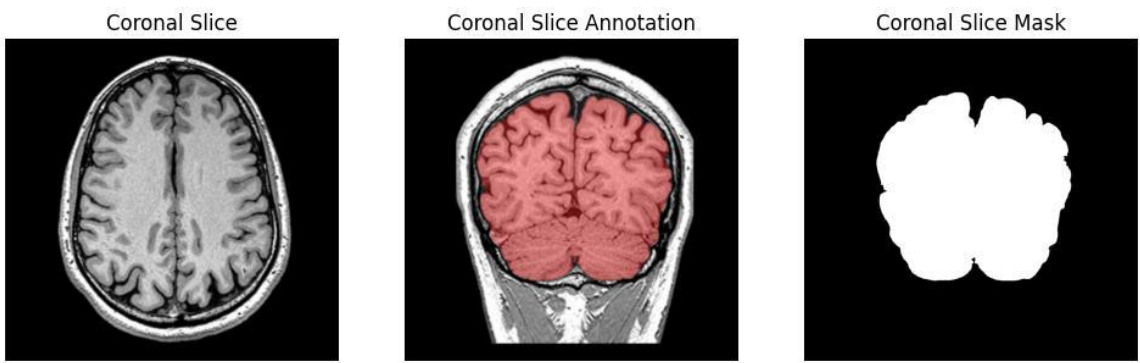

**Fig 4.** Results of brain extraction (a) MRI coronal slice (b) brain segment annotation (c) brain segmentation mask generated using SAM

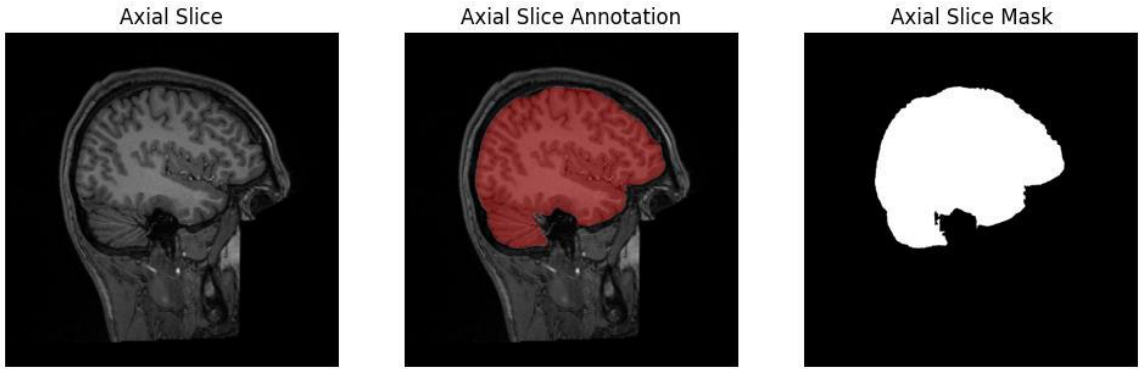

**Fig 5.** Results of brain extraction (a) MRI axial slice (b) brain segment annotation (c) brain segmentation mask generated using SAM

We compare results of SAM with manual verification by creating manual masks for brain segmentation. We use Intersection over Union (IoU) as a performance metric for measuring the correctness of the segmentation based brain extraction algorithm. The IoU score is defined as the ratio of the intersection of the predicted and ground truth regions to the union of these regions. Higher IoU values indicate better performance. The IoU score ranges from 0 to 1, where 0 means no overlap, and 1 means perfect overlap.

IoU = (Area of Intersection) / (Area of Union)

The following table presents the average IoU scores obtained from comparing SAM's segmented brain masks with the manual masks for all 30 brain MRI images:

**Table 1.** Segmentation metrics IoU across different image slices

| MRI | IoU |
|---|---|
| Sagittal slice | 0.855 |
| Coronal slice | 0.819 |
| Axial slice | 0.832 |

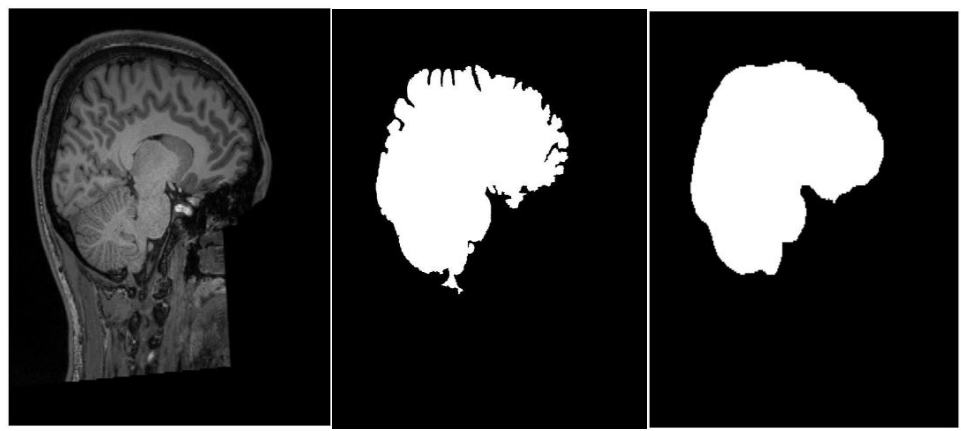

**Fig 6.** Comparing MRI sagittal image manual (left) and SAM (right) based segmentation mask

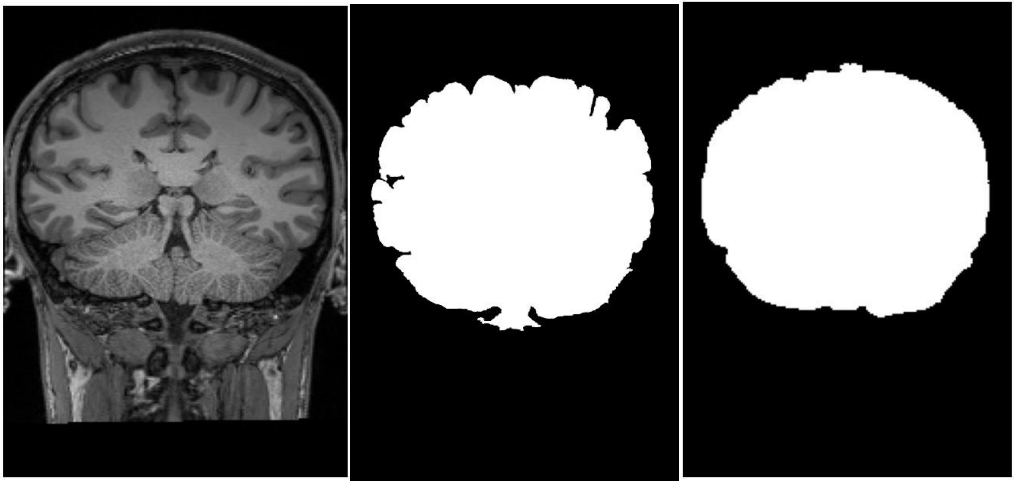

**Fig 7.** Comparing MRI sagittal image manual and SAM based segmentation mask

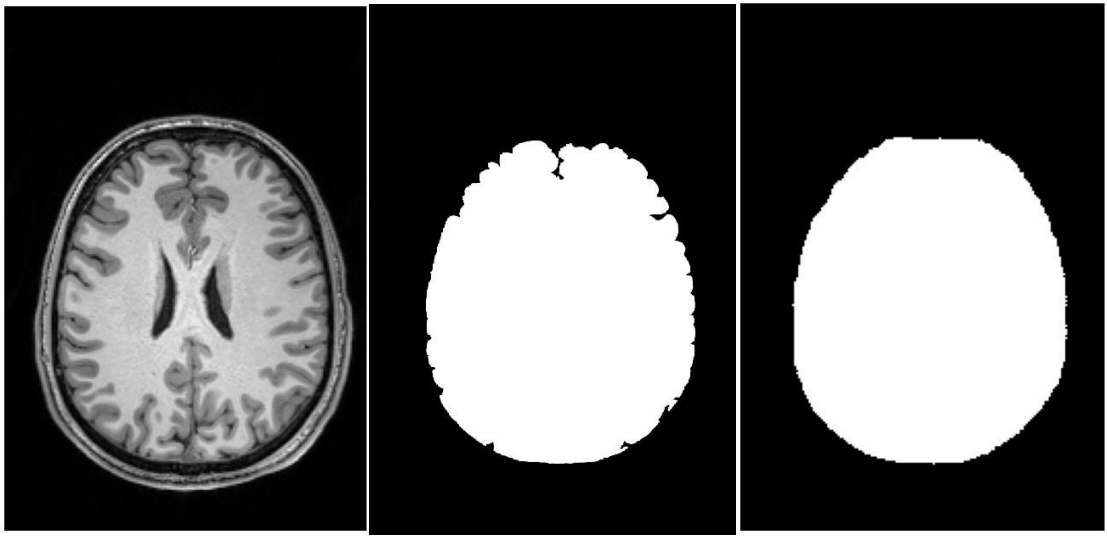

**Fig 8.** Comparing MRI sagittal image manual and SAM based segmentation mask

Table shows average metrics of 30 image slices we segmented using SAM, extracted correctly from the brain image with a IoU score of ~0.82. While there is room to improve, SAM shows great promise to be used for brain image segmentation. The results indicate a strong performance of the Segmentation-based Brain Extraction Algorithm (SAM) in brain segmentation. The IoU scores range between 0.87 and 0.92, indicating substantial overlap between SAM's segmented brain masks and the manual masks created by experts. Higher IoU scores suggest that SAM accurately identifies the brain regions, making it a reliable tool for automated brain extraction.

However, it is essential to acknowledge that manual segmentation by experts is considered the gold standard for brain extraction. While SAM demonstrates impressive accuracy, some variations in results are expected due to the complexity and variability of brain structures across different individuals and imaging modalities.

Dwith Chenna and Suyash Bhogawar## 5. CONCLUSION

In this paper we show some promising results using the Segment Anything Model (SAM), which is a generalized segmentation model, for automated brain extraction. The application of such a general segmentation model helps these neuroimaging algorithms to leverage the latest advancement, without the restriction of training the model on private medical imaging data. The results affirm the potential of SAM as a valuable tool for automated brain extraction, which can expedite the analysis of large-scale neuroimaging datasets, clinical studies, and research in the field of neuroscience. Nonetheless, further validation on diverse datasets and with larger sample sizes is recommended to establish the generalizability of SAM in various real-world scenarios.

## REFERENCES

[1] Tsang O, Gholipou A, Kehtarnavaz N, Gopinath K, Briggs R, Panahi I. Comparison of tissue segmentation algorithms in neuroimage analysis software tools. In: Proceedings of the 30th International Conference of the IEEE in Engineering in Medicine and Biology Society 2008;3924–3928.

[2] Valverde, S., Oliver, A., Cabezas, M., Roura, E., Lladó, X., 2015. Comparison of 10 brain tissue segmentation methods using revisited IBSR annotations. Journal of Magnetic Resonance Imaging 41, 93- 101.10.1002/jmri.24517

[3] S. Bonte, I. Goethals, and R. H. Van, ''Machine learning based brain Tumour segmentation on limited data using local texture and abnormality,'' Comput. Biol. Med., vol. 98, pp. 39–47, 2018.

[4] Alexander Kirillov, Eric Mintun, Nikhila Ravi, Hanzi Mao, Chloe Rolland, Laura Gustafson, Tete Xiao, Spencer Whitehead, Alexander C Berg, Wan-Yen Lo, et al. 2023. Segment Anything. arXiv preprint arXiv:2304.02643 (2023).

[5] Evans, A. C., Kamber, M., Collins, D. L., and Macdonald, D. 1994. An MRI-based probabilistic atlas of neuroanatomy. In Magnetic Resonance Scanning and Epilepsy (S. Shorvon, D. Fish, F. Andermann, G. M. Bydder, and H. Stefan, Eds.), NATO ASI Series A, Life Sciences, Vol. 264. pp. 263–274. Plenum, New York.

[6] Van Essen, D. C. et al. An integrated software suite for surface-based analyses of cerebral cortex. J. Am. Med. Inform. Assoc. 8, 443–459 (2001).

[7] Automatic anatomical brain MRI segmentation combining label propagation and decision fusion

[8] Geert L, Thijs K, Babak EB, Arnaud AAS, Francesco C, Mohsen G, Jeroen AWM, van Bram G, Clara IS. A survey on deep learning in medical image analysis. Med Image Anal. 2017;42:60–88.

[9] Alexey Dosovitskiy, Lucas Beyer, Alexander Kolesnikov, Dirk Weissenborn, Xiaohua Zhai, Thomas Unterthiner, Mostafa Dehghani, Matthias Minderer, Georg Heigold, Sylvain Gelly, Jakob Uszkoreit, and Neil Houlsby. An image is worth 16x16 words: Transformers for image recognition at scale. ICLR, 2021. 5, 8, 16

[10] Yanghao Li, Hanzi Mao, Ross Girshick, and Kaiming He. Exploring plain vision transformer backbones for object detection. ECCV, 2022. 5, 10, 11, 16, 21, 23, 24

[11] Alec Radford, Jong Wook Kim, Chris Hallacy, Aditya Ramesh, Gabriel Goh, Sandhini Agarwal, Girish Sastry, Amanda Askell, Pamela Mishkin, Jack Clark, et al. Learning transferable visual models from natural language supervision. ICML, 2021. 1, 2, 4, 5, 8, 12, 16, 22https://iaeme.com/Home/journal/IJAIMED 7 editor@iaeme.com